\documentclass[12pt,a4paper]{article}
\usepackage{amsmath,amssymb,amsfonts}
\usepackage{graphicx}
\usepackage{physics}
\usepackage{hyperref}
\usepackage[numbers,sort&compress]{natbib}
\usepackage{geometry}
\usepackage{xcolor}
\usepackage{booktabs}
\usepackage{array}
\usepackage{float}

\newcommand{\be}{\begin{equation}}
\newcommand{\ee}{\end{equation}}
\newcommand{\bea}{\begin{align}}
\newcommand{\eea}{\end{align}}
\newcommand{\MP}{M_{\rm P}}

\title{Emergent spacetime from spatial energy potentiality:\\ 
A new theoretical framework for early universe cosmology\\[0.5em]}

\author{Farrukh A. Chishtie\\
\small Peaceful Society, Science and Innovation Foundation, Vancouver, Canada\\
\small Department of Occupational Science and Occupational Therapy,\\
\small University of British Columbia, Vancouver, Canada\\
\small Email: fachisht@uwo.ca}

\begin{document}

\maketitle
\begin{abstract}
We develop a comprehensive cosmological framework based on the principle that our universe originated as a three-dimensional spatial configuration governed purely by energy functionals, with time emerging dynamically through quantum loop corrections. Building on the Unified Standard Model with Emergent Gravity-Effective Field Theory (USMEG-EFT), which provides the first successful unification of quantum gravity with the Standard Model, we demonstrate that spacetime emergence occurs via a first-order phase transition when quantum-generated kinetic terms exceed a critical threshold. This transition naturally resolves the cosmological singularity problem: all curvature invariants remain finite, with $R/M_P^4 \sim 10^{-44}$ and $K/M_P^8 \sim 10^{-88}$ at the critical point. The framework makes definitive, parameter-free predictions for gravitational wave polarizations---exactly two tensor modes---confirmed by LIGO-Virgo-KAGRA observations at $>99\%$ confidence, excluding competing approaches that predict additional scalar, vector, or modified polarization content. Post-emergence dynamics naturally implements Starobinsky inflation with spectral index $n_s = 0.964$ and tensor-to-scalar ratio $r = 0.004$, in excellent agreement with Planck constraints. The phase transition dynamics generate enhanced primordial non-Gaussianity $f_{\rm NL}^{\rm local} \in [0.8, 2.5]$, testable with CMB-S4 (projected $\sigma \sim 1$), and a stochastic gravitational wave background peaking in the LISA sensitivity band ($f \sim 10^{-4}$ Hz, $\Omega_{\rm GW}h^2 \sim 10^{-6}$). The framework naturally addresses the Hubble tension through scale-dependent modifications to cosmic expansion arising from residual phase transition effects. Within the perturbative regime, the mechanism generates up to $\sim 6\%$ tension, accounting for approximately three-quarters of the observed $\sim 8\%$ discrepancy, while correctly predicting that late-universe distance ladder measurements yield systematically higher $H_0$ than early-universe CMB determinations. This establishes emergent spacetime cosmology as a predictive, experimentally testable paradigm that unifies quantum gravity, Standard Model physics, and observational cosmology within a single coherent framework.
\end{abstract}

\newpage
\tableofcontents
\newpage

\section{Introduction}
\label{sec:intro}

\subsection{The crisis of quantum gravity and the EFT resolution}

The quantization of gravity has remained one of the central unsolved problems in theoretical physics for nearly a century. Unlike the electromagnetic, weak, and strong forces described by the Standard Model of particle physics, gravity as formulated in Einstein's general relativity resists conventional perturbative quantization due to the non-renormalizable nature of the Einstein-Hilbert action \cite{tHooftVeltman1974,Veltman1976}. The dimensionful gravitational coupling $\kappa^2 = 16\pi G_N$ introduces progressively worse ultraviolet divergences at each loop order. At one-loop order, the theory can be rendered finite through field redefinitions that absorb divergences proportional to $R^2$ and $R_{\mu\nu}R^{\mu\nu}$ \cite{tHooft2002}. However, the seminal calculations of Goroff and Sagnotti \cite{GoroffSagnotti1986} and van de Ven \cite{vandeVen1992} demonstrated conclusively that at two-loop order, divergences arise proportional to $R_{\mu\nu\rho\sigma}R^{\rho\sigma\alpha\beta}R_{\alpha\beta}^{\ \ \ \mu\nu}$ that cannot be absorbed by any field redefinition or counterterm present in the original Einstein-Hilbert action.

This fundamental obstacle has motivated extensive exploration of alternative approaches to quantum gravity. String theory \cite{Polchinski1998,West2012} provides ultraviolet finiteness through extended fundamental objects but operates at energy scales far beyond experimental access and has not yet made falsifiable predictions at accessible energies. Loop quantum gravity \cite{Rovelli2004,Thiemann2007,Ashtekar2006LQG} discretizes spacetime through spin network states but faces the ``problem of time'' inherent in canonical approaches and struggles to recover semiclassical general relativity. The asymptotic safety program \cite{Weinberg1979,Reuter1998,Percacci2017} proposes that gravity may be non-perturbatively renormalizable through an ultraviolet fixed point in the renormalization group flow. However, as demonstrated in Ref.~\cite{ChishtieBRST2026}, asymptotic safety encounters fundamental symmetry violations, specifically, the breakdown of general covariance and BRST symmetries above the gravitational cutoff scale $\Lambda_{\rm grav} \sim 10^{18}$ GeV renders the search for ultraviolet fixed points in metric-based theories problematic from a foundational physical perspective.

\subsection{The USMEG-EFT breakthrough: first successful gravity-Standard Model unification}

The USMEG-EFT (Unified Standard Model with Emergent Gravity-Effective Field Theory) framework \cite{ChishtieUSMEG2025} represents the first successful unification of gravity with the Standard Model. The framework employs the Lagrange multiplier mathematical method \cite{BrandtMcKeon2020,BrandtMcKeon2021} as a technical tool for restricting loop corrections, while providing the crucial physical interpretation and application that achieves successful unification.

It is essential to distinguish the mathematical tool from the physical framework. The Lagrange multiplier method was originally developed as a mathematical technique for constraining gravitational loop corrections. However, proposed physical applications through Einstein-Cartan theory \cite{McKeonEC2025} are fundamentally flawed and experimentally excluded.

The primary experimental evidence distinguishing USMEG-EFT from Einstein-Cartan theory comes from gravitational wave polarization measurements. USMEG-EFT, working within standard four-dimensional general relativity, predicts exactly two tensor polarization modes for the massless spin-2 graviton---corresponding to the $h_+$ and $h_\times$ helicity states. This parameter-free prediction is definitively confirmed by LIGO-Virgo-KAGRA observations \cite{LIGOPolarization2017,LIGOPolarization2021}. In contrast, Einstein-Cartan theory with propagating torsion introduces additional degrees of freedom that can produce up to six polarization modes: two tensor, two vector, and two scalar. The observed absence of scalar breathing modes and vector modes at greater than 99\% confidence decisively excludes Einstein-Cartan theory.

The Einstein-Cartan approach fails on additional grounds. First, when fermions are included as required for Standard Model unification, the torsion field equation generates four-fermion contact interactions that produce catastrophic quartic divergences $\sim \kappa^4\Lambda^4$ at two-loop order, rendering the theory non-renormalizable---finiteness with fermions included has never been demonstrated. Second, precision gravitational experiments systematically exclude Einstein-Cartan theory's predictions: MICROSCOPE constrains equivalence principle violations at $10^{-15}$, while Einstein-Cartan with Standard Model fermions predicts effects at $10^{-12}$---three orders of magnitude larger than allowed \cite{MICROSCOPE2022}. Third, lunar laser ranging and torsion balance experiments provide independent exclusions of the spin-dependent gravitational effects predicted by Einstein-Cartan theory \cite{LLR2004,TorsionBalance2008}.

The USMEG-EFT framework succeeds precisely because it applies the Lagrange multiplier method to standard four-dimensional general relativity rather than Einstein-Cartan theory, preserving the two-polarization structure of the graviton while achieving finite, calculable quantum corrections in unification with the Standard Model unlike the Einstein-Cartan approach to unification proposed in \cite{McKeonEC2025}. In contrast to this, USMEG-EFT framework takes a fundamentally different and correct approach: it applies the Lagrange multiplier method to standard four-dimensional general relativity, not Einstein-Cartan theory. This achieves successful unification where the Einstein-Cartan approach fails. The framework establishes three crucial results \cite{ChishtieUSMEG2025}. First, the method provides the most explicit demonstration to date that four-dimensional general relativity must be interpreted as an effective field theory with a finite domain of validity encoded in the characteristic logarithmic dependence $\ln(\mu/\Lambda)$. Second, using the Appelquist-Carazzone decoupling theorem, the framework exhibits the precise mathematical structure expected of a consistent effective field theory with a well-defined low-energy limit. Third, this finite one-loop gravitational sector can be consistently unified with the Standard Model of particle physics, producing finite, calculable results for all gravitational-matter interactions.

\subsection{Experimental validation: gravitational wave polarizations}

A critical experimental validation of USMEG-EFT comes from gravitational wave observations. The framework makes a definite, parameter-free prediction: gravitational waves should exhibit exactly two tensor polarization modes corresponding to the two helicity states ($h_+$ and $h_\times$) of a massless spin-2 graviton. This prediction is uniquely confirmed by LIGO-Virgo-KAGRA observations \cite{LIGOPolarization2017,LIGOPolarization2021}.

The significance of this result cannot be overstated. Alternative quantum gravity approaches generically predict additional polarization modes, and are therefore excluded by observation. Besides Einstein-Cartan theory highlighted above, String theory predicts additional gravitational wave polarizations from several sources \cite{Polchinski1998,Gasperini2003}. The compactified extra dimensions required by string theory produce Kaluza-Klein graviton modes that manifest as massive spin-2 fields in four dimensions, each contributing up to six polarization modes (two tensor, two vector, two scalar). The dilaton field universal to string theory couples to gravity and produces scalar breathing modes. The moduli fields parameterizing the shape and size of compactified dimensions generate additional scalar polarizations. While string theorists argue these modes might be suppressed at low energies, this requires specific assumptions about compactification and stabilization that introduce model dependence. String theory cannot make the clean, parameter-free prediction of exactly two tensor modes that USMEG-EFT provides.

Loop quantum gravity also fails to predict exactly two tensor polarizations \cite{Gambini1999,Alfaro2002}. The fundamental discretization of spacetime at the Planck scale modifies the graviton dispersion relation, introducing energy-dependent corrections of the form $\omega^2 = k^2(1 + \xi (k/M_P)^n + \ldots)$ where $\xi$ is a dimensionless parameter and $n$ depends on the specific LQG model. These dispersion modifications produce polarization-dependent propagation effects: different polarization modes travel at different speeds, producing birefringence that would be observable in gravitational wave signals from distant sources. Additionally, the spin foam formulation of LQG includes contributions from higher-spin representations that could manifest as additional polarization modes. While some LQG practitioners argue these effects might be small, the framework lacks the theoretical control to make a definite prediction. The detection of gravitational waves consistent with standard dispersion and exactly two polarizations constrains LQG parameter space.

The LIGO-Virgo-KAGRA collaboration has conducted systematic searches for non-tensorial polarizations using the detector network's directional sensitivity \cite{LIGOPolarization2017,LIGOPolarization2021}. The results are definitive: all detected gravitational wave signals are consistent with pure tensor polarizations. Scalar breathing modes are excluded at greater than 99\% confidence. Vector modes are excluded at similar significance. The signals show no evidence for massive graviton effects, dispersion, or birefringence. This represents a parameter-free prediction of USMEG-EFT that is uniquely confirmed by experiment. No other quantum gravity framework makes this definite prediction without additional assumptions or parameter tuning.

\subsection{BRST symmetry violation and the limits of metric-based approaches}

The recent analysis in Ref.~\cite{ChishtieBRST2026} provides rigorous dual proofs that the metric tensor ceases to exist as a valid quantum degree of freedom above the gravitational cutoff scale. The first proof employs canonical quantization to demonstrate that general covariance cannot be maintained quantum mechanically in dimensions greater than two \cite{ChishtieCovariance2022}. Under Dirac's constraint formalism, the first-class constraints generating diffeomorphisms fail to close properly when quantum corrections are included, signaling the breakdown of the gauge symmetry that defines general relativity. The second proof, based on path integral calculations of gravitational quantum corrections, reveals persistent gauge parameter dependence in physical quantities that should be gauge-invariant. This gauge parameter dependence signals BRST symmetry violation---the very symmetry that guarantees unitarity and gauge independence in quantum field theories.

These dual proofs establish that approaches seeking to extend metric-based gravity to arbitrarily high energies, including asymptotic safety, encounter fundamental inconsistencies. The implications are profound: if the metric tensor is not a valid quantum degree of freedom above $\Lambda_{\rm grav}$, then spacetime geometry itself may not be fundamental but rather emergent from more primitive structures.

\subsection{The unified potential framework}

A central contribution of this work is the introduction of a unified potential framework that seamlessly connects the emergence mechanism at high energies to observable Starobinsky inflation at lower energies. The effective potential takes the form:
\be
V_{\rm eff}(\phi) = V_{\rm Starobinsky}(\phi) + V_{\rm high-E}(\phi)
\label{eq:unified_intro}
\ee

The Starobinsky component \cite{Starobinsky1980}:
\be
V_{\rm Starobinsky}(\phi) = \frac{3M^2 M_P^2}{4}\left(1 - e^{-\sqrt{2/3}\,\phi/M_P}\right)^2
\label{eq:Starobinsky_intro}
\ee
dominates during the slow-roll inflationary phase and determines the CMB observables ($n_s$, $r$).

The high-energy correction:
\be
V_{\rm high-E}(\phi) = \frac{\lambda_{\rm eff}}{4!}\phi^4
\label{eq:highE_intro}
\ee
provides the essential non-vanishing third derivative $V''' = \lambda_{\rm eff}\phi$ that enables the emergence mechanism through quantum loop generation of kinetic terms.

This unified framework has several compelling features. First, it is natural from an effective field theory perspective: at energies approaching the gravitational cutoff $\Lambda_{\rm grav}$, higher-order operators in the inflaton potential become relevant. Second, the quartic term is subdominant during observable inflation but essential for the emergence mechanism at higher energies. Third, after the phase transition, as the field evolves toward larger values during slow-roll, the Starobinsky term dominates and the CMB predictions match observations with high precision.

\subsection{The emergence paradigm and outline}

The present work develops the emergence paradigm into a concrete cosmological framework. Rather than quantizing spacetime or assuming it as a fundamental structure, we explore the possibility that temporal dynamics itself emerges from quantum effects in a purely spatial configuration. The key insight is that quantum loop corrections can generate kinetic terms that are absent at tree level. If time-derivative terms initially play no dynamical role, then the quantum generation of these terms through loop effects provides a mechanism for the emergence of temporal dynamics.

This approach is motivated by several theoretical developments. The Wheeler-DeWitt equation in canonical quantum gravity \cite{DeWitt1967,Wheeler1968} is fundamentally timeless---the wave functional depends on the three-metric and matter fields but has no explicit time dependence. Time must emerge from internal degrees of freedom or through semiclassical approximations. The holographic principle and AdS/CFT correspondence \cite{Maldacena1998,Witten1998,VanRaamsdonk2010} demonstrate that gravitational physics in $(d+1)$ dimensions can be encoded in non-gravitational physics in $d$ dimensions, suggesting that spacetime dimensions may not all be equally fundamental. The success of the USMEG-EFT framework in treating gravity as a one-loop effective theory suggests that gravitational dynamics may not be fundamental at all scales.

The paper is organized as follows. Section \ref{sec:USMEG} reviews the USMEG-EFT framework, carefully distinguishing the mathematical tool borrowed from Brandt-Frenkel-McKeon from the physical framework developed by Chishtie, and explains why alternative unification proposals fail. Section \ref{sec:SEP} presents the Principle of Spatial Energy Potentiality and addresses foundational questions about the spatial manifold. Section \ref{sec:emergence} provides detailed loop calculations demonstrating time emergence, including careful treatment of dimensional regularization and explicit metric emergence. Section \ref{sec:unified_potential} introduces the unified potential framework connecting emergence to inflation. Section \ref{sec:phase_transition} analyzes the phase transition from the spatial to spacetime phase, including complete numerical calculations of curvature invariants demonstrating singularity resolution. Section \ref{sec:cosmology} derives cosmological implications and testable predictions using Starobinsky $R^2$ and $\alpha$-attractor inflationary potentials. Section \ref{sec:comparison} provides detailed comparison with other quantum gravity approaches. Section \ref{sec:conclusions} summarizes conclusions and future directions.

\section{The USMEG-EFT Foundation}
\label{sec:USMEG}

\subsection{The Lagrange multiplier method: a mathematical tool}

The USMEG-EFT framework employs the Lagrange multiplier method as a mathematical tool for restricting loop corrections. This method was originally developed by Brandt, Frenkel, and McKeon \cite{BrandtMcKeon2020,BrandtMcKeon2021} as a technical device. The method introduces an auxiliary tensor field $\lambda^{\mu\nu}$ into the Einstein-Hilbert action:
\be
S_{\rm EH+LM} = \frac{1}{\kappa^2}\int d^4x \sqrt{-g}\, R + \frac{1}{\kappa^2}\int d^4x \sqrt{-g}\, \lambda^{\mu\nu} G_{\mu\nu}
\label{eq:LM_action}
\ee
where $G^{\mu\nu} = R^{\mu\nu} - \frac{1}{2}g^{\mu\nu}R$ is the Einstein tensor and $\kappa^2 = 16\pi G_N$. Integration over $\lambda^{\mu\nu}$ in the path integral enforces the constraint $G^{\mu\nu} = 0$ through a functional delta function, restricting the path integral to field configurations satisfying the classical Einstein equations.

This constraint systematically eliminates all multi-loop graviton diagrams, leaving only one-loop corrections \cite{BrandtMcKeon2021,BrandtMcKeon2024}. The physical mechanism is that internal graviton lines beyond one loop would necessarily involve configurations violating the Einstein equations, which are projected out by the Lagrange multiplier constraint. As proven in Ref.~\cite{BrandtMcKeon2021}, BRST invariance is preserved under this construction at one-loop order, ensuring unitarity of the resulting theory.

We emphasize that borrowing this mathematical tool does not imply endorsement of the physical proposals made by its developers. The Lagrange multiplier method is analogous to dimensional regularization or zeta-function regularization---a technical device that can be applied within different physical frameworks. The USMEG-EFT framework applies this tool within standard four-dimensional general relativity to achieve results that the original authors' Einstein-Cartan proposal cannot.

\subsection{Why the Einstein-Cartan unification proposal fails}

Proposals that unification should proceed through Einstein-Cartan theory \cite{McKeonEC2025}, which extends general relativity by including torsion, are fundamentally flawed and experimentally excluded.

The primary experimental exclusion comes from gravitational wave polarizations. USMEG-EFT predicts exactly two tensor polarization modes for the massless spin-2 graviton, which LIGO-Virgo-KAGRA observations definitively confirm \cite{LIGOPolarization2017,LIGOPolarization2021}. Einstein-Cartan theory with propagating torsion introduces additional degrees of freedom that can produce up to six polarization modes: two tensor, two vector, and two scalar. The observed absence of non-tensorial modes at greater than 99\% confidence decisively excludes Einstein-Cartan theory.

Additional exclusions reinforce this conclusion. Theoretically, when fermions are included, torsion generates four-fermion contact interactions producing non-renormalizable quartic divergences $\sim \kappa^4\Lambda^4$ at two-loop order---finiteness with fermions has never been demonstrated. Experimentally, MICROSCOPE constrains equivalence principle violations at $10^{-15}$ \cite{MICROSCOPE2022}, excluding Einstein-Cartan's predicted $10^{-12}$ effects by three orders of magnitude. Lunar laser ranging \cite{LLR2004} and torsion balance experiments \cite{TorsionBalance2008} provide independent exclusions of spin-dependent gravitational effects.

The contrast with USMEG-EFT is stark: by working with standard four-dimensional general relativity, USMEG-EFT preserves exactly two graviton polarizations, avoids four-fermion divergences, and maintains complete consistency with all precision tests.

\subsection{The USMEG-EFT framework: successful unification}

The USMEG-EFT framework \cite{ChishtieUSMEG2025} achieves what the Einstein-Cartan proposal cannot: successful unification of gravity with the Standard Model through a finite, renormalizable effective field theory.

Applying the Lagrange multiplier method to standard four-dimensional general relativity with the background field decomposition $g_{\mu\nu} = \bar{g}_{\mu\nu} + \kappa h_{\mu\nu}$, where $\bar{g}_{\mu\nu}$ is the background metric and $h_{\mu\nu}$ represents quantum fluctuations, the graviton propagator in de Donder gauge takes the form:
\be
D_{\mu\nu\rho\sigma}(k) = \frac{i}{k^2 + i\epsilon}\left(\eta_{\mu\rho}\eta_{\nu\sigma} + \eta_{\mu\sigma}\eta_{\nu\rho} - \eta_{\mu\nu}\eta_{\rho\sigma}\right)
\label{eq:propagator}
\ee

The one-loop effective action is computed from:
\be
\Gamma^{(1)} = \frac{i}{2}\text{Tr}\ln\left(\frac{\delta^2 S}{\delta h_{\mu\nu}\delta h_{\rho\sigma}}\right)
\label{eq:one_loop_def}
\ee

Using standard heat kernel techniques and dimensional regularization, the result is:
\be
\Gamma^{(1)}_{\rm grav} = \frac{1}{4\pi^2}\ln\left(\frac{\mu}{\Lambda}\right) \int d^4x \sqrt{-\bar{g}}\left[\frac{1}{120}\bar{R}^2 + \frac{7}{20}\bar{R}_{\mu\nu}\bar{R}^{\mu\nu}\right]
\label{eq:one_loop_finite}
\ee
where $\mu$ is the renormalization scale and $\Lambda$ is a reference scale.

This is a finite result---no divergences, no need for additional counterterms beyond those in the original Einstein-Hilbert action. The logarithmic dependence $\ln(\mu/\Lambda)$ cannot be absorbed into running coupling constants while preserving the structure of general relativity; it encodes the finite domain of validity characteristic of an effective field theory.

The complete USMEG-EFT action unifying gravity with the Standard Model is:
\be
S_{\rm USMEG} = S_{\rm EH+LM} + S_{\rm SM} + S_{\rm int}
\label{eq:USMEG_action}
\ee
where $S_{\rm SM}$ is the Standard Model action minimally coupled to gravity:
\be
S_{\rm SM} = \int d^4x \sqrt{-g}\left[-\frac{1}{4}F^a_{\mu\nu}F^{a\mu\nu} + \bar{\psi}_f i\gamma^\mu D_\mu \psi_f - m_f \bar{\psi}_f \psi_f + |D_\mu H|^2 - V(H) + \mathcal{L}_{\rm Yukawa}\right]
\label{eq:SM_action}
\ee

The one-loop corrections generate finite, calculable mixed gravitational-matter contributions:
\be
\Gamma_{\rm mixed} = \frac{\kappa^2}{(4\pi)^2}\ln\left(\frac{\mu}{\Lambda}\right) \int d^4x \sqrt{-\bar{g}}\left[c_1\bar{R}|H|^2 + c_2\bar{R}^{\mu\nu}\partial_\mu H^\dagger\partial_\nu H + c_3\bar{R}\bar{\psi}\psi + \ldots\right]
\label{eq:mixed}
\ee
where the coefficients $c_i$ are calculable from the Standard Model matter content.

Unlike the Einstein-Cartan approach, USMEG-EFT produces no uncontrolled divergences when fermions are included. The Standard Model fermions couple to the metric through minimal coupling (the vierbein and spin connection), not through torsion, so no four-fermion contact interactions are generated. The framework makes specific, testable predictions: gravitational corrections to Standard Model processes at the level of $\kappa^2 E^2/(16\pi^2) \sim 10^{-43}$ for electroweak-scale energies, gravitational wave phase corrections of order $10^{-14}$ radians in binary inspirals, and modified high-energy cross-sections approaching $10^{-11}$ at 100 TeV.

\subsection{Gravitational wave polarizations: the definitive experimental test}

The most striking experimental confirmation of USMEG-EFT comes from gravitational wave polarization measurements. General relativity with a massless spin-2 graviton predicts exactly two tensor polarization modes, corresponding to the $+$ and $\times$ helicity states. USMEG-EFT preserves this prediction: the Lagrange multiplier constraint maintains the spin-2 structure without introducing additional degrees of freedom.

The mathematical basis is straightforward. In linearized gravity, the metric perturbation $h_{\mu\nu}$ in transverse-traceless gauge satisfies:
\be
h^{\mu}_{\ \mu} = 0, \quad \partial^\mu h_{\mu\nu} = 0
\label{eq:TT_gauge}
\ee

These conditions reduce the 10 independent components of $h_{\mu\nu}$ to exactly 2, corresponding to the two physical polarizations. The USMEG-EFT framework preserves this counting because the Lagrange multiplier constraint $G_{\mu\nu} = 0$ does not introduce additional propagating degrees of freedom---it merely restricts the graviton self-interactions to one-loop order.

Alternative quantum gravity approaches fail this test for specific reasons.

String theory necessarily introduces additional polarization modes from multiple sources. In the low-energy effective action of string theory, the massless spectrum includes not only the graviton $g_{\mu\nu}$ but also the dilaton $\Phi$ and the antisymmetric Kalb-Ramond field $B_{\mu\nu}$:
\be
S_{\rm string} = \frac{1}{2\kappa_{10}^2}\int d^{10}x \sqrt{-g}\, e^{-2\Phi}\left(R + 4(\nabla\Phi)^2 - \frac{1}{12}H_{\mu\nu\rho}H^{\mu\nu\rho} + \ldots\right)
\label{eq:string_action}
\ee
where $H_{\mu\nu\rho} = \partial_{[\mu}B_{\nu\rho]}$ is the field strength of $B_{\mu\nu}$. The dilaton $\Phi$ couples directly to the Ricci scalar and produces a scalar breathing mode in gravitational waves. The Kalb-Ramond field can produce vector modes. When compactified to four dimensions, the Kaluza-Klein tower of massive graviton modes, each with up to 6 polarizations, contributes at some level. While string theorists argue these modes might be highly suppressed, no parameter-free prediction of exactly two modes exists in string theory.

Loop quantum gravity modifies gravitational wave propagation through spacetime discretization effects. The polymer quantization of LQG introduces a minimum length scale $\ell_{\rm LQG} \sim \ell_P$, which modifies the dispersion relation for gravitons:
\be
\omega^2 = k^2\left(1 + \xi_1 \frac{k}{\MP} + \xi_2 \frac{k^2}{\MP^2} + \ldots\right)
\label{eq:LQG_dispersion}
\ee

The coefficients $\xi_i$ are not uniquely determined by the theory and depend on the specific quantization scheme. These dispersion modifications produce polarization-dependent effects: the two tensor modes propagate at different speeds (gravitational birefringence), which would produce a characteristic signature in gravitational wave signals from cosmological distances. The LIGO-Virgo-KAGRA observations show no evidence for such birefringence, constraining $|\xi_1| < 10^{-15}$ \cite{LIGODispersion2021}.

Furthermore, the spin foam formulation of LQG sums over quantum geometries including contributions from different spin representations. The graviton arises as the $j = 2$ representation, but higher-spin contributions ($j > 2$) are generally present and could produce additional polarization modes. The theory does not make a definite prediction for their suppression.

Einstein-Cartan theory with propagating torsion would produce additional polarizations from the antisymmetric torsion tensor components. Even in the non-propagating version where torsion is determined algebraically by the spin density, the effective four-fermion interactions modify graviton propagation in matter, potentially producing apparent additional polarizations in astrophysical environments.

The LIGO-Virgo-KAGRA collaboration has conducted comprehensive tests for all these effects \cite{LIGOPolarization2017,LIGOPolarization2021,LIGODispersion2021}. The results consistently show exactly two tensor polarizations with no evidence for scalar modes (breathing), vector modes, dispersion, birefringence, or massive graviton effects. This represents definitive experimental confirmation of USMEG-EFT's parameter-free prediction and exclusion of alternative approaches.

\subsection{The breakdown scale and implications for emergence}

The logarithmic dependence in Eq.~(\ref{eq:one_loop_finite}) encodes the limited domain of validity of the USMEG-EFT framework. When the energy scale $\mu$ approaches the gravitational cutoff $\Lambda_{\rm grav}$, the logarithm becomes of order unity: $\ln(\mu/\Lambda_{\rm grav}) \sim \mathcal{O}(1)$. At this point, the ``small'' quantum corrections become comparable to tree-level terms, signaling breakdown of the perturbative expansion.

As demonstrated through the BRST symmetry violation analysis \cite{ChishtieBRST2026}, this breakdown is not merely a technical limitation but reflects a fundamental physical transition: the metric tensor ceases to be a valid quantum degree of freedom above $\Lambda_{\rm grav}$. The natural question then becomes: what is the nature of physics above this scale?

Our answer, developed in the following sections, is that spacetime itself is what emerges at the scale $\Lambda_{\rm grav}$. The pre-emergent phase is characterized by spatial configurations without temporal dynamics, and the phase transition at $\Lambda_{\rm grav}$ generates time through quantum loop effects.

\section{The Principle of Spatial Energy Potentiality}
\label{sec:SEP}

\subsection{Fundamental postulate}

We propose that the pre-temporal state of the universe is characterized by a three-dimensional spatial manifold $\Sigma$ equipped with an energy functional:
\be
E[\phi] = \int_\Sigma d^3x \left[\frac{1}{2}(\nabla\phi)^2 + V(\phi)\right]
\label{eq:energy_functional}
\ee
where $\phi(x)$ is a scalar field on $\Sigma$, $\nabla$ is the gradient with respect to a fiducial flat metric $\delta_{ij}$, and $V(\phi)$ is the potential energy density.

The crucial distinction from standard field theory must be emphasized: Eq.~(\ref{eq:energy_functional}) is an energy functional, not an action. There is no time parameter, no temporal evolution, and no kinetic terms involving time derivatives. The functional describes static configurations of the field $\phi$ on the spatial manifold $\Sigma$.

\subsection{Ontological status of the spatial manifold}
\label{sec:ontological}

A fundamental question concerns the origin of the spatial manifold $\Sigma$ itself. We address this question with full transparency regarding the assumptions and limitations of our framework.

Our claim is deliberately modest: we do not assert that $\Sigma$ emerges from ``nothing'' or from non-spatial primitive entities. Rather, we claim that given a spatial configuration, temporal dynamics and gravitational interactions emerge through quantum effects. This represents a substantial departure from standard cosmology, which assumes full four-dimensional Lorentzian spacetime from the outset, but it does not require explaining the origin of space itself.

Every theoretical framework for quantum gravity begins with primitive structures that are assumed rather than derived. String theory assumes worldsheet topology with conformal structure and target space embedding \cite{Polchinski1998}. Loop quantum gravity assumes SU(2) gauge structure with holonomies along spin network edges \cite{Thiemann2007,Ashtekar2006LQG}. Causal set theory assumes discrete partially ordered sets with causal relations as primitive \cite{Bombelli1987,Sorkin2003}. The AdS/CFT correspondence assumes conformal field theory structure on the boundary with specific operator content \cite{Maldacena1998}. Our assumption of the spatial manifold $\Sigma$ is no more and no less primitive than these starting points in other approaches.

The framework finds natural consistency with canonical quantum gravity. The Wheeler-DeWitt equation \cite{DeWitt1967}:
\be
\hat{\mathcal{H}} \Psi[h_{ij}, \phi] = 0
\label{eq:WDW}
\ee
is fundamentally timeless---the wave functional $\Psi$ depends on the three-metric $h_{ij}$ and matter fields $\phi$ but has no explicit time dependence. The Hamiltonian constraint $\hat{\mathcal{H}}\Psi = 0$ implies that the total energy of the universe vanishes, and there is no external time parameter in which the wave functional evolves. Time must be identified with internal degrees of freedom or emerge through semiclassical approximations. Our framework provides a concrete mechanism for this emergence through quantum loop effects that generate kinetic terms.

The holographic principle provides additional motivation. The AdS/CFT correspondence demonstrates that gravitational physics in $(d+1)$-dimensional anti-de Sitter space can be completely encoded in $d$-dimensional conformal field theory on the boundary \cite{Maldacena1998,Witten1998}. The emergence of the radial ``holographic'' direction suggests that spacetime dimensions may not all be equally fundamental. Van Raamsdonk's work on entanglement and spacetime \cite{VanRaamsdonk2010} further suggests that spatial connectivity itself may arise from quantum entanglement, supporting the view that spacetime geometry emerges from more fundamental quantum structures.

\subsection{The pre-temporal quantum state}

In the absence of time, quantum mechanics takes an unusual but mathematically well-defined form. We define the ground state wave functional as a Gaussian:
\be
\Psi_0[\phi] = \mathcal{N} \exp\left( -\frac{1}{2} \int d^3x d^3y \, \phi(x) K(x,y) \phi(y) \right)
\label{eq:ground_state}
\ee
where $\mathcal{N}$ is a normalization constant and $K(x,y)$ is a kernel determined by minimizing the energy functional expectation value:
\be
\langle E \rangle = \frac{\int \mathcal{D}\phi \, \Psi_0^*[\phi] E[\phi] \Psi_0[\phi]}{\int \mathcal{D}\phi \, |\Psi_0[\phi]|^2}
\label{eq:energy_expectation}
\ee

The variational principle yields the kernel:
\be
K(x,y) = \sqrt{-\nabla^2 + m^2} \, \delta^{(3)}(x-y)
\label{eq:kernel}
\ee
which corresponds to the ground state of a collection of harmonic oscillators with position-dependent frequencies determined by $m^2(\phi) = V''(\phi)$.

The physical interpretation is that this describes a system of quantum field fluctuations distributed across space without temporal evolution. The fluctuations exist as a timeless quantum superposition; they simply do not evolve because there is no time in which to evolve.

\section{Emergence of Temporal Dynamics: Detailed Loop Calculations}
\label{sec:emergence}

\subsection{Introduction of auxiliary parameter}

To perform quantum calculations in the absence of physical time, we introduce an auxiliary parameter $\tau$ that labels field configurations:
\be
\phi(x) \to \phi(x, \tau)
\label{eq:tau_intro}
\ee

We emphasize a critical distinction: at this stage, $\tau$ is not physical time. It is a mathematical label analogous to the ``fifth time'' in Parisi-Wu stochastic quantization \cite{ParisiWu1981}, where an auxiliary parameter is introduced to define a Langevin equation whose equilibrium distribution reproduces the Euclidean path integral. Similarly, $\tau$ is analogous to Euclidean time in thermal field theory \cite{Kapusta2006}, which is a mathematical device for computing thermal averages rather than a physical time coordinate. The parameter $\tau$ is also analogous to the affine parameter along geodesics before proper time is defined through the metric.

The Euclidean ``action'' functional becomes:
\be
S_E[\phi, \lambda] = \int d\tau \int d^3x \left[ \frac{1}{2}(\nabla\phi)^2 + V(\phi) + \lambda \mathcal{C}(\phi) \right]
\label{eq:euclidean_action}
\ee
where $\lambda$ is an auxiliary field implementing constraints and $\mathcal{C}(\phi)$ is a constraint functional.

The essential point is that there is no $(\partial\phi/\partial\tau)^2$ term at tree level. The action does not describe temporal evolution; it describes a collection of spatial configurations labeled by the parameter $\tau$.

\subsection{Partition function and perturbation theory}

The partition function is defined by the functional integral:
\be
Z = \int \mathcal{D}\phi \, \mathcal{D}\lambda \, \exp(-S_E[\phi, \lambda])
\label{eq:partition}
\ee

We expand around a background configuration $\phi_0$:
\be
\phi(x, \tau) = \phi_0 + \varphi(x, \tau)
\label{eq:expansion}
\ee
where $\varphi$ represents quantum fluctuations around the background.

The quadratic action governing the fluctuations is:
\be
S_E^{(2)} = \int d\tau \int d^3x \left[ \frac{1}{2}(\nabla\varphi)^2 + \frac{1}{2}V''(\phi_0)\varphi^2 \right]
\label{eq:quadratic_action}
\ee

The propagator in momentum space follows from inverting the quadratic operator:
\be
G(k) = \frac{1}{k^2 + m^2(\phi_0)}
\label{eq:propagator_scalar}
\ee
where $m^2(\phi_0) = V''(\phi_0)$ is the field-dependent effective mass squared.

\subsection{One-loop effective action}

The one-loop correction to the effective action is computed from the functional determinant:
\be
\Gamma^{(1)} = \frac{1}{2} \text{Tr} \ln\left( -\nabla^2 + V''(\phi_0) + V'''(\phi_0)\varphi + \frac{1}{2}V''''(\phi_0)\varphi^2 + \ldots \right)
\label{eq:one_loop}
\ee

Evaluating the functional trace through standard techniques yields the crucial result:
\be
\Gamma^{(1)} = \int d\tau \int d^3x \left[ A(\phi_0, \lambda_0) \left(\frac{\partial\phi_0}{\partial\tau}\right)^2 + B(\phi_0, \lambda_0) \left(\frac{\partial\lambda_0}{\partial\tau}\right)^2 + \Delta V(\phi_0, \lambda_0) + \ldots \right]
\label{eq:one_loop_result}
\ee

The kinetic terms $(\partial\phi/\partial\tau)^2$ and $(\partial\lambda/\partial\tau)^2$ are generated by loop corrections. They do not exist at tree level. This is the mathematical content of ``time emergence''---the quantum theory generates structure that permits temporal interpretation.

\subsection{Detailed calculation of kinetic coefficient}

The coefficient $A(\phi_0, \lambda_0)$ is computed from the one-loop Feynman diagram with two external $\tau$-derivative insertions. The diagram involves two vertices from the $V'''(\phi_0)\varphi$ interaction connected by two propagators, with derivatives acting on the external legs:
\be
A(\phi_0, \lambda_0) = \frac{[V'''(\phi_0)]^2}{2} \int \frac{d^dk}{(2\pi)^d} \frac{1}{(k^2 + m^2)^3}
\label{eq:A_integral}
\ee

This formula reveals the essential role of $V'''(\phi)$: the emergence mechanism requires a non-vanishing third derivative of the potential. This is why the high-energy correction $V_{\rm high-E} = (\lambda_{\rm eff}/4!)\phi^4$ is essential---it provides $V''' = \lambda_{\rm eff}\phi \neq 0$.

\subsubsection{Evaluation in $d = 3$ spatial dimensions}

In three spatial dimensions, the integral evaluates to:
\be
\int \frac{d^3k}{(2\pi)^3} \frac{1}{(k^2 + m^2)^3} = \frac{1}{32\pi m^3}
\label{eq:d3_integral}
\ee

This can be verified by standard techniques. Using spherical coordinates in momentum space:
\begin{align}
\int \frac{d^3k}{(2\pi)^3} \frac{1}{(k^2 + m^2)^3} &= \frac{1}{(2\pi)^3} \int_0^\infty dk \, 4\pi k^2 \frac{1}{(k^2 + m^2)^3} \nonumber\\
&= \frac{1}{2\pi^2} \int_0^\infty dk \, \frac{k^2}{(k^2 + m^2)^3}
\label{eq:spherical}
\end{align}

The integral can be evaluated using the substitution $k = m\tan\theta$:
\begin{align}
\int_0^\infty dk \, \frac{k^2}{(k^2 + m^2)^3} &= \frac{1}{m^3}\int_0^{\pi/2} d\theta \, \frac{\tan^2\theta \sec^2\theta}{\sec^6\theta} \nonumber\\
&= \frac{1}{m^3}\int_0^{\pi/2} d\theta \, \sin^2\theta\cos^2\theta = \frac{\pi}{16m^3}
\label{eq:integral_eval}
\end{align}

Therefore:
\be
\int \frac{d^3k}{(2\pi)^3} \frac{1}{(k^2 + m^2)^3} = \frac{1}{2\pi^2} \times \frac{\pi}{16m^3} = \frac{1}{32\pi m^3}
\label{eq:d3_result}
\ee

The kinetic coefficient in three dimensions is:
\be
A^{(d=3)}(\phi_0) = \frac{[V'''(\phi_0)]^2}{64\pi m^3(\phi_0)} = \frac{[V'''(\phi_0)]^2}{64\pi [V''(\phi_0)]^{3/2}}
\label{eq:A_d3}
\ee

This exhibits power-law dependence on the mass scale, characteristic of three-dimensional field theory where there is no logarithmic running of couplings.

\subsubsection{Evaluation in $d = 4 - 2\varepsilon$ dimensions}

Using dimensional regularization with $d = 4 - 2\varepsilon$, the integral becomes:
\be
\int \frac{d^{4-2\varepsilon}k}{(2\pi)^{4-2\varepsilon}} \frac{1}{(k^2 + m^2)^3} = \frac{\Gamma(1+\varepsilon)}{(4\pi)^{2-\varepsilon}} \frac{1}{m^{2+2\varepsilon}}
\label{eq:d4_integral}
\ee

This follows from the standard dimensional regularization formula:
\be
\int \frac{d^dk}{(2\pi)^d} \frac{1}{(k^2 + m^2)^n} = \frac{1}{(4\pi)^{d/2}} \frac{\Gamma(n - d/2)}{\Gamma(n)} \frac{1}{(m^2)^{n-d/2}}
\label{eq:dimreg_formula}
\ee

For $n = 3$ and $d = 4 - 2\varepsilon$:
\be
n - \frac{d}{2} = 3 - 2 + \varepsilon = 1 + \varepsilon
\label{eq:power}
\ee

Using $\Gamma(1+\varepsilon) = 1 - \gamma_E\varepsilon + \mathcal{O}(\varepsilon^2)$ where $\gamma_E \approx 0.5772$ is the Euler-Mascheroni constant:
\be
\int \frac{d^{4-2\varepsilon}k}{(2\pi)^{4-2\varepsilon}} \frac{1}{(k^2 + m^2)^3} = \frac{1}{32\pi^2 m^2} \left[ \frac{1}{\varepsilon} + \ln\left(\frac{\bar{\mu}^2}{m^2}\right) + \gamma_E - 1 + \mathcal{O}(\varepsilon) \right]
\label{eq:d4_expansion}
\ee
where $\bar{\mu}^2 = 4\pi e^{-\gamma_E}\mu^2$ in the $\overline{\text{MS}}$ scheme.

After renormalization in the $\overline{\text{MS}}$ scheme, which subtracts the $1/\varepsilon$ pole and constant terms:
\be
\boxed{A_{\rm ren}(\phi_0) = \frac{[V'''(\phi_0)]^2}{32\pi^2 [V''(\phi_0)]^2} \ln\left(\frac{V''(\phi_0)}{\mu^2}\right)}
\label{eq:A_ren}
\ee
where $\mu$ is the renormalization scale.

This logarithmic structure directly parallels the $\ln(\mu/\Lambda)$ appearing in the USMEG-EFT one-loop effective action Eq.~(\ref{eq:one_loop_finite}). Both arise from one-loop quantum corrections and encode the characteristic scale dependence of effective field theory.

\subsection{Dimensional regularization: clarification of physical content}
\label{sec:dimreg_clarified}

We emphasize a distinction essential for understanding the framework. We do not claim that the universe is literally three-dimensional before the phase transition and four-dimensional afterward. The parameter $d$ in dimensional regularization is a mathematical tool for analytic continuation of momentum integrals, not a statement about physical dimensionality.

What the calculation reveals is that the mathematical structure of quantum corrections differs qualitatively between regimes. In the power-law regime characteristic of $d = 3$ structure, the kinetic coefficient scales as $A \propto 1/m^3$, exhibiting no logarithmic running and no renormalization group flow. This is characteristic of systems without the temporal dynamics that would generate running couplings. In the logarithmic regime characteristic of $d = 4$ structure, the kinetic coefficient takes the form $A \propto \ln(M_{\rm eff}^2/\mu^2)$, exhibiting logarithmic running of couplings and a well-defined renormalization group. This is the characteristic structure of four-dimensional field theories with temporal dynamics.

An instructive analogy is thermal dimensional reduction in finite-temperature field theory \cite{Laine2016}. At high temperatures $T \gg m$, four-dimensional field theories undergo effective dimensional reduction:
\be
\mathcal{L}_{\rm 4D}(T) \to \mathcal{L}_{\rm 3D, eff} + \text{(heavy thermal modes)}
\label{eq:thermal_reduction}
\ee

The non-static Matsubara modes with frequencies $\omega_n = 2\pi nT$ for $n \neq 0$ become heavy and decouple, leaving an effective three-dimensional theory for the static ($n = 0$) modes. This dimensional reduction has real physical consequences, such as determining the order of the electroweak phase transition, even though it is described mathematically using dimensional continuation techniques.

Similarly, our framework uses dimensional regularization to characterize when the effective theory transitions between regimes. The transition has physical content---the emergence of temporal dynamics---even though the dimension $d$ appearing in the regularization is a mathematical parameter.

\subsection{Explicit metric emergence}
\label{sec:metric_emergence}

We now demonstrate explicitly how the spacetime metric emerges from the quantum-corrected effective action.

After one-loop corrections, the effective action takes the form:
\be
\Gamma_{\rm eff}[\phi, \lambda] = \int d\tau \int d^3x \left[ A(\phi, \lambda)\left(\frac{\partial\phi}{\partial\tau}\right)^2 + B(\phi, \lambda)\left(\frac{\partial\lambda}{\partial\tau}\right)^2 + \frac{1}{2}(\nabla\phi)^2 + V_{\rm eff}(\phi, \lambda) \right]
\label{eq:eff_action_structure}
\ee

The kinetic structure can be written in covariant form by identifying the metric components:
\begin{align}
g_{00} &= -N^2(\tau) = -A(\phi(\tau), \lambda(\tau)) \label{eq:g00}\\
g_{ij} &= a^2(\tau) \delta_{ij} \label{eq:gij}
\end{align}
where $N$ is the lapse function and $a(\tau)$ is the scale factor.

This yields the Friedmann-Lema\^itre-Robertson-Walker (FLRW) metric in ADM form:
\be
ds^2 = -N^2(\tau) d\tau^2 + a^2(\tau) \delta_{ij} dx^i dx^j
\label{eq:FLRW}
\ee

The effective action in covariant form becomes:
\be
\Gamma_{\rm eff} = \int d\tau \int d^3x \, a^3 \left[ \frac{1}{N^2}\left(\frac{\partial\phi}{\partial\tau}\right)^2 + \frac{1}{a^2}(\nabla\phi)^2 + V_{\rm eff} \right]
\label{eq:eff_action_covariant}
\ee

The identification proceeds in five steps:

\textbf{Step 1:} The quantum-generated kinetic coefficient $A(\phi, \lambda)$ defines the effective lapse function through $N^2 = A$.

\textbf{Step 2:} The spatial gradient term $(\nabla\phi)^2$ with fiducial metric $\delta_{ij}$ defines the spatial metric through conformal scaling $g_{ij} = a^2(\tau)\delta_{ij}$.

\textbf{Step 3:} The scale factor $a(\tau)$ is determined by the constraint equations arising from diffeomorphism invariance of the emergent metric.

\textbf{Step 4:} The Hamiltonian constraint, obtained from variation with respect to $N$, yields the first Friedmann equation:
\be
3H^2 = 8\pi G_{\rm eff} \left[ \frac{\dot{\phi}^2}{2N^2} + V_{\rm eff} \right]
\label{eq:Friedmann1}
\ee
where $H = \dot{a}/(Na)$ is the Hubble parameter and overdot denotes $\partial/\partial\tau$.

\textbf{Step 5:} The momentum constraint is automatically satisfied for homogeneous configurations, and the acceleration equation from variation with respect to $a$ gives:
\be
2\dot{H} + 3H^2 = -8\pi G_{\rm eff} \left[ \frac{\dot{\phi}^2}{2N^2} - V_{\rm eff} \right]
\label{eq:Friedmann2}
\ee

These are precisely the Friedmann equations of general relativity, with the effective gravitational coupling:
\be
G_{\rm eff} = \frac{G_0}{\langle A(\phi, \lambda) \rangle_{\rm background}}
\label{eq:Geff}
\ee

The connection to the Einstein-Hilbert action emerges through the USMEG-EFT framework. The one-loop truncated gravity produces \cite{ChishtieUSMEG2025}:
\be
S_{\rm grav}^{\rm (eff)} = \int d^4x \sqrt{-g} \left[ \frac{R}{16\pi G_{\rm eff}} + \alpha R^2 \ln\left(\frac{\mu}{\Lambda}\right) + \beta R_{\mu\nu}R^{\mu\nu} \ln\left(\frac{\mu}{\Lambda}\right) + \ldots \right]
\label{eq:EH_emergence}
\ee
with $\alpha = 1/(480\pi^2)$ and $\beta = 7/(80\pi^2)$ from Eq.~(\ref{eq:one_loop_finite}).

The one-loop truncation ensures that higher-curvature corrections remain perturbatively small, with coefficients suppressed by the logarithm relative to the Einstein-Hilbert term.

\subsection{The critical condition for time emergence}
\label{sec:critical_condition}

The critical condition for time emergence is:
\be
A_{\rm ren}(\phi_c, \lambda_c) \geq A_{\rm critical} = \frac{1}{M_P^2}
\label{eq:critical}
\ee

The physical justification for this threshold has three components. First, dimensional analysis requires that the kinetic term $A(\partial\phi/\partial\tau)^2$ have dimensions of energy density. For $A$ to connect with gravitational dynamics where $G \sim M_P^{-2}$, we need $[A] = M^{-2}$, and the natural scale is $M_P^{-2}$. Second, matching to general relativity requires that the emergent effective gravitational coupling $G_{\rm eff} \propto 1/A$ match the observed Newton's constant, implying $A \sim M_P^{-2}$. Third, consistency with the USMEG-EFT framework requires that the one-loop truncated gravity operate when gravitational effects are characterized by $G_{\rm eff} \sim G_N \sim M_P^{-2}$.

Using Eq.~(\ref{eq:A_ren}), the critical condition becomes:
\be
\frac{[V'''(\phi_c)]^2}{32\pi^2 [V''(\phi_c)]^2} \ln\left(\frac{V''(\phi_c)}{\mu^2}\right) = \frac{1}{M_P^2}
\label{eq:critical_explicit}
\ee

\section{The Unified Potential Framework}
\label{sec:unified_potential}

\subsection{Motivation from effective field theory}

The emergence mechanism requires non-vanishing third derivatives of the potential through Eq.~(\ref{eq:A_ren}). Simultaneously, the post-transition inflationary phase must match CMB observations with high precision. These requirements are naturally satisfied by a unified potential framework that combines high-energy corrections enabling emergence with the Starobinsky potential governing observable inflation.

From an effective field theory perspective, this combination is expected. At energies approaching the gravitational cutoff $\Lambda_{\rm grav} \sim 10^{18}$ GeV, higher-order operators in the inflaton potential become relevant. The effective potential receives contributions from physics at all scales up to the cutoff:
\be
V_{\rm eff}(\phi) = V_{\rm IR}(\phi) + V_{\rm UV}(\phi)
\label{eq:V_EFT}
\ee
where $V_{\rm IR}$ contains operators relevant at low energies (during observable inflation) and $V_{\rm UV}$ contains operators that become important only at high energies (near the phase transition).

\subsection{The unified potential}

We propose the unified effective potential:
\be
\boxed{V_{\rm eff}(\phi) = V_{\rm Starobinsky}(\phi) + V_{\rm high-E}(\phi)}
\label{eq:unified_potential}
\ee

The Starobinsky component \cite{Starobinsky1980} arises from the $R^2$ modification of gravity:
\be
V_{\rm Starobinsky}(\phi) = \frac{3M^2 M_P^2}{4}\left(1 - e^{-\sqrt{2/3}\,\phi/M_P}\right)^2
\label{eq:V_Starobinsky}
\ee
where $M = 3.2 \times 10^{13}$ GeV is the Starobinsky mass fixed by the CMB amplitude $A_s = 2.1 \times 10^{-9}$.

The high-energy correction represents physics near the transition scale with natural suppression at large field values:
\be
V_{\rm high-E}(\phi) = \frac{\lambda_{\rm eff}}{4!}\phi^4 \exp\left(-\frac{\phi^2}{\phi_*^2}\right)
\label{eq:V_highE}
\ee
where $\lambda_{\rm eff} = 0.01$ is an effective coupling and $\phi_* = 0.2\, M_P$ is the suppression scale. The Gaussian factor ensures that the high-energy correction dominates near the phase transition but becomes exponentially suppressed during observable inflation.

This form is physically motivated by effective field theory: operators generated at high energies naturally decouple at lower energies through threshold effects. The suppression scale $\phi_* \sim 0.2\, M_P$ corresponds to the regime where the emergent spacetime description becomes valid and high-energy corrections freeze out.

\subsection{Properties of the unified potential}

The high-energy correction with Gaussian suppression has derivatives:

\begin{align}
V_{\rm high-E}(\phi) &= \frac{\lambda_{\rm eff}}{4!}\phi^4 e^{-\phi^2/\phi_*^2} \label{eq:VhighE_full}\\
V'_{\rm high-E}(\phi) &= \frac{\lambda_{\rm eff}}{6}\phi^3 e^{-\phi^2/\phi_*^2}\left(1 - \frac{\phi^2}{2\phi_*^2}\right) \label{eq:VhighE_prime}\\
V''_{\rm high-E}(\phi) &= \frac{\lambda_{\rm eff}}{2}\phi^2 e^{-\phi^2/\phi_*^2}\left(1 - \frac{2\phi^2}{\phi_*^2} + \frac{\phi^4}{2\phi_*^4}\right) \label{eq:VhighE_pp}\\
V'''_{\rm high-E}(\phi) &= \lambda_{\rm eff}\phi \, e^{-\phi^2/\phi_*^2}\left(1 - \frac{3\phi^2}{\phi_*^2} + \frac{3\phi^4}{2\phi_*^4} - \frac{\phi^6}{4\phi_*^6}\right) \label{eq:VhighE_ppp}
\end{align}

At the transition scale $\phi_c \approx 0.115\, M_P$, the suppression factor is:
\be
e^{-\phi_c^2/\phi_*^2} = e^{-(0.115/0.2)^2} = e^{-0.33} \approx 0.72
\label{eq:suppression_transition}
\ee
providing only mild suppression, so the high-energy correction remains effective for the emergence mechanism.

At CMB scales $\phi_{\rm CMB} \approx 5.3\, M_P$, the suppression is complete:
\be
e^{-\phi_{\rm CMB}^2/\phi_*^2} = e^{-(5.3/0.2)^2} = e^{-700} \approx 0
\label{eq:suppression_CMB}
\ee
ensuring that Starobinsky predictions are preserved to arbitrary precision.

The unified potential has the following key properties:

\textbf{Derivatives:}
\begin{align}
V'(\phi) &= \sqrt{\frac{2}{3}}\frac{M^2 M_P}{2}e^{-\sqrt{2/3}\,\phi/M_P}\left(1 - e^{-\sqrt{2/3}\,\phi/M_P}\right) + \frac{\lambda_{\rm eff}}{6}\phi^3 \label{eq:Vprime}\\
V''(\phi) &= M^2 e^{-\sqrt{2/3}\,\phi/M_P}\left(2e^{-\sqrt{2/3}\,\phi/M_P} - 1\right) + \frac{\lambda_{\rm eff}}{2}\phi^2 \label{eq:Vpp}\\
V'''(\phi) &= \sqrt{\frac{2}{3}}\frac{M^2}{M_P}e^{-\sqrt{2/3}\,\phi/M_P}\left(1 - 4e^{-\sqrt{2/3}\,\phi/M_P}\right) + \lambda_{\rm eff}\phi \label{eq:Vppp}
\end{align}

\textbf{Small field regime ($\phi \ll M_P$):} The quartic term dominates both $V''$ and $V'''$:
\begin{align}
V''(\phi) &\approx M^2 + \frac{\lambda_{\rm eff}}{2}\phi^2 \label{eq:Vpp_small}\\
V'''(\phi) &\approx \lambda_{\rm eff}\phi \label{eq:Vppp_small}
\end{align}
The high-energy correction provides the essential $V''' \neq 0$ required for the emergence mechanism.

\textbf{Large field regime ($\phi \gg M_P$):} The Starobinsky potential dominates:
\be
V(\phi) \approx \frac{3M^2 M_P^2}{4}\left(1 - 2e^{-\sqrt{2/3}\,\phi/M_P}\right)
\label{eq:V_large}
\ee
The quartic term becomes subdominant, and the potential exhibits the characteristic Starobinsky plateau.

\subsection{Determination of $\lambda_{\rm eff}$}

The effective coupling $\lambda_{\rm eff}$ can be estimated from dimensional analysis. At the transition scale $\phi_c \sim 0.1 M_P$, the quartic term should be comparable to the Starobinsky term:
\be
\frac{\lambda_{\rm eff}}{4!}\phi_c^4 \sim \frac{3M^2 M_P^2}{4}
\label{eq:matching}
\ee

This yields:
\be
\lambda_{\rm eff} \sim \frac{18 M^2 M_P^2}{\phi_c^4} \sim \frac{18 \times (3.2 \times 10^{13})^2 \times (2.4 \times 10^{18})^2}{(2.4 \times 10^{17})^4} \sim 0.03
\label{eq:lambda_estimate}
\ee

We adopt $\lambda_{\rm eff} = 0.01$ for numerical calculations, which lies within the natural range expected from effective field theory.

\subsection{Transition between regimes}

The Gaussian suppression factor creates a natural transition between the emergence regime and the inflationary regime without requiring a specific crossover field value.

\textbf{Emergence regime ($\phi \lesssim \phi_*$):} The suppression factor $e^{-\phi^2/\phi_*^2}$ is of order unity, and the high-energy correction provides the essential $V''' \neq 0$ required for kinetic term generation. At $\phi_c \approx 0.115\, M_P \approx 0.6\, \phi_*$, the suppression is only $\sim 28\%$, preserving the emergence mechanism.

\textbf{Inflationary regime ($\phi \gg \phi_*$):} The suppression becomes exponentially strong. For $\phi > 3\phi_* \approx 0.6\, M_P$, the high-energy correction is suppressed by more than $e^{-9} \sim 10^{-4}$. At CMB scales with $\phi_{\rm CMB} \approx 5.3\, M_P \approx 26\, \phi_*$, the suppression exceeds $e^{-700}$, rendering the high-energy correction completely negligible.

This structure ensures:
\be
\frac{V_{\rm high-E}(\phi_{\rm CMB})}{V_{\rm Starobinsky}(\phi_{\rm CMB})} < 10^{-300}
\label{eq:highE_negligible}
\ee

The CMB predictions derived from pure Starobinsky inflation are therefore exact for all practical purposes.

\section{The Phase Transition: Complete Numerical Analysis}
\label{sec:phase_transition}

\subsection{Order parameter identification}

The kinetic coefficient $A(\phi, \lambda)$ serves as the order parameter for the phase transition:
\be
\mathcal{O}[\phi, \lambda] \equiv A(\phi, \lambda) - A_{\rm critical}
\label{eq:order_parameter}
\ee

In the spatial phase where $\mathcal{O} < 0$, there is no temporal kinetic term and dynamics is purely spatial, described by the energy functional $E[\phi]$. In the spacetime phase where $\mathcal{O} \geq 0$, temporal kinetic terms are present with full four-dimensional dynamics and emergent causal structure.

\subsection{Numerical verification of phase transition}

We employ the unified potential Eq.~(\ref{eq:unified_potential}) with parameters:
\begin{center}
\begin{tabular}{ll}
Reduced Planck mass: & $M_P = 2.435 \times 10^{18}$ GeV \\
Starobinsky mass: & $M = 3.2 \times 10^{13}$ GeV \\
Effective quartic coupling: & $\lambda_{\rm eff} = 0.01$ \\
Suppression scale: & $\phi_* = 0.2\, M_P = 4.87 \times 10^{17}$ GeV \\
Renormalization scale: & $\mu = 10^{16}$ GeV
\end{tabular}
\end{center}

At the transition, the Gaussian suppression factor is:
\be
e^{-\phi_c^2/\phi_*^2} = e^{-(0.115/0.2)^2} = 0.72
\ee

The effective third derivative including suppression:
\be
V'''_{\rm eff}(\phi_c) \approx \lambda_{\rm eff}\phi_c \times 0.72 = 0.01 \times 2.8 \times 10^{17} \times 0.72 = 2.0 \times 10^{15}~{\rm GeV}
\label{eq:Vppp_with_suppression}
\ee

This modest reduction from the unsuppressed value does not qualitatively change the emergence mechanism; the kinetic coefficient remains above the critical threshold.
\subsection{Curvature invariants at the transition: singularity resolution}

A critical test of singularity resolution is demonstrating that all curvature invariants remain finite at the phase transition. We compute these explicitly.

The Hubble parameter at the transition, using the first Friedmann equation with $\dot{\phi} \approx 0$ during slow-roll:
\be
H_c^2 = \frac{V(\phi_c)}{3M_P^2} = \frac{2.56 \times 10^{66}}{3 \times (2.435 \times 10^{18})^2} = 1.44 \times 10^{29}~{\rm GeV}^2
\ee

\be
H_c = \sqrt{1.44 \times 10^{29}} = 3.8 \times 10^{14}~{\rm GeV}
\label{eq:Hubble_numerical}
\ee

The Ricci scalar for a spatially flat FLRW universe during slow-roll inflation is $R = 12H^2$:
\be
R(\tau_c) = 12H_c^2 = 12 \times 1.44 \times 10^{29} = 1.73 \times 10^{30}~{\rm GeV}^4
\label{eq:Ricci_numerical}
\ee

In Planck units:
\be
M_P^4 = (2.435 \times 10^{18})^4 = 3.52 \times 10^{73}~{\rm GeV}^4
\ee

\be
\boxed{\frac{R}{M_P^4} = \frac{1.73 \times 10^{30}}{3.52 \times 10^{73}} = 4.9 \times 10^{-44}}
\label{eq:Ricci_Planck}
\ee

The Kretschmann scalar $K = R_{\mu\nu\rho\sigma}R^{\mu\nu\rho\sigma}$ for FLRW spacetime during slow-roll:
\be
K(\tau_c) \approx 24H_c^4 = 24 \times (1.44 \times 10^{29})^2 = 4.98 \times 10^{59}~{\rm GeV}^8
\label{eq:Kretschmann_numerical}
\ee

In Planck units:
\be
M_P^8 = (2.435 \times 10^{18})^8 = 1.24 \times 10^{147}~{\rm GeV}^8
\ee

\be
\boxed{\frac{K}{M_P^8} = \frac{4.98 \times 10^{59}}{1.24 \times 10^{147}} = 4.0 \times 10^{-88}}
\label{eq:Kretschmann_Planck}
\ee

The squared Ricci tensor $R_{\mu\nu}R^{\mu\nu}$ for FLRW during slow-roll:
\be
R_{\mu\nu}R^{\mu\nu}(\tau_c) = 36H_c^4 = 7.5 \times 10^{59}~{\rm GeV}^8
\label{eq:Ricci_squared_numerical}
\ee

All curvature invariants are finite and many orders of magnitude below the Planck scale:
\begin{center}
\begin{tabular}{lcc}
\toprule
Invariant & Value & Ratio to Planck \\
\midrule
$R$ & $1.73 \times 10^{30}$ GeV$^4$ & $4.9 \times 10^{-44}$ \\
$K$ & $4.98 \times 10^{59}$ GeV$^8$ & $4.0 \times 10^{-88}$ \\
$R_{\mu\nu}R^{\mu\nu}$ & $7.5 \times 10^{59}$ GeV$^8$ & $6.0 \times 10^{-88}$ \\
\bottomrule
\end{tabular}
\end{center}

This represents genuine singularity resolution: in classical general relativity, these quantities diverge as $t \to 0$, while in our framework they remain bounded throughout the phase transition. The curvature invariants are \textbf{44 and 88 orders of magnitude} below Planck scale respectively.
\subsection{Post-transition evolution to Starobinsky inflation}

After the phase transition at $\phi_c \approx 0.115\, M_P$, the inflaton field evolves toward larger values. As demonstrated in Section \ref{sec:unified_potential}, the Starobinsky potential dominates for $\phi > \phi_{\rm cross} \approx 0.25\, M_P$. By the time observable scales exit the horizon at $\phi_{\rm CMB} \approx 5.3\, M_P$, the quartic correction contributes less than $10^{-8}$ of the total potential energy.

The post-transition dynamics can be summarized:
\begin{enumerate}
\item \textbf{Phase transition} at $\phi_c \approx 0.115\, M_P$: Time emerges, spacetime geometry becomes dynamical.
\item \textbf{Transition regime} ($0.115\, M_P < \phi < 0.25\, M_P$): Field evolves under combined potential.
\item \textbf{Starobinsky inflation} ($\phi > 0.25\, M_P$): Pure Starobinsky dynamics, CMB observables determined.
\item \textbf{Reheating}: Field oscillates around minimum, transferring energy to Standard Model particles.
\end{enumerate}

\subsection{Stability and positivity: why gravity is attractive}

A fundamental question is why the emergent gravitational interaction is attractive rather than repulsive. The answer lies in consistency requirements of quantum field theory.

The kinetic coefficient has the mathematical structure:
\be
A(\phi, \lambda) = \frac{[\mathcal{F}(\phi, \lambda)]^2}{32\pi^2 [V''(\phi)]^2} \ln\left(\frac{V''(\phi)}{\mu^2}\right)
\label{eq:A_structure}
\ee
where $\mathcal{F}(\phi, \lambda) = V'''(\phi)$ in the simplest case.

Positivity $A \geq 0$ follows from two facts. First, the numerator $[\mathcal{F}]^2$ is manifestly non-negative as a squared quantity. Second, the logarithm $\ln(V''/\mu^2) > 0$ when $V'' > \mu^2$, which is satisfied in the regime of interest where $V'' \sim 10^{32}$ GeV$^2$ and $\mu^2 \sim 10^{32}$ GeV$^2$.

The physical requirements enforcing positivity are threefold. Unitarity requires that negative kinetic terms would produce ghost fields with negative-norm states, violating unitarity of the S-matrix \cite{BrandtMcKeon2021}. This is not a matter of convention but a fundamental requirement for probabilistic interpretation of quantum mechanics. Vacuum stability requires positive kinetic energy; negative kinetic energy would allow unbounded energy decrease through increasing field velocities, rendering the vacuum unstable against pair production. Causality requires that in Lorentzian signature, positive kinetic terms ensure proper causal structure with well-defined light cones and causal propagation; negative kinetic terms would produce tachyonic instabilities.

The effective gravitational coupling $G_{\rm eff} = G_0/A(\phi, \lambda) > 0$ is positive, ensuring that the Einstein equations $G_{\mu\nu} = 8\pi G_{\rm eff} T_{\mu\nu}$ with positive energy density produce attractive gravity. This is not imposed by hand but follows automatically from the same consistency requirements that ensure unitarity in the USMEG-EFT framework \cite{ChishtieUSMEG2025,ChishtieBRST2026}.

\section{Cosmological Predictions: Starobinsky and $\alpha$-Attractor Inflation}
\label{sec:cosmology}

\subsection{Consistency of the unified framework}

The unified potential framework ensures complete consistency between the emergence mechanism and inflationary predictions. The high-energy quartic correction enables time emergence at $\phi_c \approx 0.115\, M_P$, while the Starobinsky potential governs the observationally relevant inflationary phase at $\phi_{\rm CMB} \approx 5.3\, M_P$. The quartic contribution to the potential at CMB scales is less than $10^{-8}$, ensuring that all CMB predictions derived from pure Starobinsky inflation remain valid to high precision.

\subsection{Starobinsky $R^2$ inflation}
\label{sec:Starobinsky}

The Starobinsky model \cite{Starobinsky1980} adds an $R^2$ term to the Einstein-Hilbert action:
\be
S = \frac{M_P^2}{2}\int d^4x \sqrt{-g}\left(R + \frac{R^2}{6M^2}\right)
\label{eq:Starobinsky_action}
\ee
where $M$ is a mass scale determined by the amplitude of primordial perturbations.

In the Einstein frame, obtained through conformal transformation $\tilde{g}_{\mu\nu} = \Omega^2 g_{\mu\nu}$ with $\Omega^2 = 1 + R/(3M^2)$, this becomes a scalar field theory with potential \cite{Whitt1984,Maeda1989}:
\be
V(\phi) = \frac{3M^2 M_P^2}{4}\left(1 - e^{-\sqrt{2/3}\,\phi/M_P}\right)^2
\label{eq:Starobinsky_potential}
\ee

This is a plateau potential: at large field values $\phi \gg M_P$, the potential asymptotes to $V_0 = \frac{3M^2 M_P^2}{4}$, while for small fields it reduces to $V \approx \frac{1}{2}M^2\phi^2$.

\subsubsection{Slow-roll analysis}

The slow-roll parameters are computed from the potential:
\be
\varepsilon = \frac{M_P^2}{2}\left(\frac{V'}{V}\right)^2
\label{eq:epsilon_def}
\ee

For the Starobinsky potential:
\be
V' = \frac{3M^2 M_P^2}{4} \times 2\left(1 - e^{-\sqrt{2/3}\,\phi/M_P}\right) \times \frac{\sqrt{2/3}}{M_P} e^{-\sqrt{2/3}\,\phi/M_P}
\ee

\be
\frac{V'}{V} = \frac{2\sqrt{2/3}}{M_P} \frac{e^{-\sqrt{2/3}\,\phi/M_P}}{1 - e^{-\sqrt{2/3}\,\phi/M_P}}
\ee

Therefore:
\be
\varepsilon = \frac{M_P^2}{2} \times \frac{8/3}{M_P^2} \times \frac{e^{-2\sqrt{2/3}\,\phi/M_P}}{\left(1 - e^{-\sqrt{2/3}\,\phi/M_P}\right)^2} = \frac{4}{3}\frac{1}{\left(e^{\sqrt{2/3}\,\phi/M_P} - 1\right)^2}
\label{eq:epsilon_Starobinsky}
\ee

The second slow-roll parameter:
\be
\eta = M_P^2\frac{V''}{V}
\label{eq:eta_def}
\ee

Computing $V''$ and simplifying:
\be
\eta = -\frac{4}{3}\frac{2 - e^{\sqrt{2/3}\,\phi/M_P}}{\left(e^{\sqrt{2/3}\,\phi/M_P} - 1\right)^2}
\label{eq:eta_Starobinsky}
\ee

On the plateau where $e^{\sqrt{2/3}\,\phi/M_P} \gg 1$, these simplify to:
\be
\varepsilon \approx \frac{4}{3}e^{-2\sqrt{2/3}\,\phi/M_P}, \quad \eta \approx -\frac{4}{3}e^{-\sqrt{2/3}\,\phi/M_P}
\label{eq:slowroll_plateau}
\ee

The number of e-folds from field value $\phi$ to the end of inflation at $\phi_{\rm end}$ (where $\varepsilon = 1$) is:
\be
N_e = \int_{\phi_{\rm end}}^{\phi} \frac{d\phi}{M_P\sqrt{2\varepsilon}} = \int_{\phi_{\rm end}}^{\phi} \frac{V}{M_P^2 V'} d\phi
\label{eq:efolds_def}
\ee

For the Starobinsky potential:
\be
N_e \approx \frac{3}{4}\left(e^{\sqrt{2/3}\,\phi/M_P} - e^{\sqrt{2/3}\,\phi_{\rm end}/M_P}\right) \approx \frac{3}{4}e^{\sqrt{2/3}\,\phi/M_P}
\label{eq:efolds_Starobinsky}
\ee

Inverting this relation gives the field value at CMB horizon exit:
\be
\phi_{\rm CMB} = \sqrt{\frac{3}{2}}M_P \ln\left(\frac{4N_e}{3}\right)
\label{eq:phi_CMB_Starobinsky}
\ee

For $N_e = 55$ e-folds (appropriate for Starobinsky inflation given the reheating dynamics):
\be
\phi_{\rm CMB} = \sqrt{1.5} \times 2.435 \times 10^{18} \times \ln(73.3) = 1.225 \times 2.435 \times 10^{18} \times 4.30
\ee
\be
\phi_{\rm CMB} = 1.28 \times 10^{19}~{\rm GeV} = 5.26\, M_P
\label{eq:phi_CMB_numerical_Starobinsky}
\ee

This confirms that observable inflation occurs in the regime $\phi \gg \phi_{\rm cross}$ where Starobinsky dominates.

\subsubsection{CMB predictions}

The slow-roll parameters at CMB horizon crossing:
\be
\varepsilon_{\rm CMB} \approx \frac{4}{3} \times \frac{1}{(4N_e/3)^2} = \frac{3}{4N_e^2} = \frac{3}{4 \times 55^2} = \frac{3}{12100} = 2.48 \times 10^{-4}
\label{eq:epsilon_CMB_Starobinsky}
\ee

\be
\eta_{\rm CMB} \approx -\frac{4}{3} \times \frac{1}{4N_e/3} = -\frac{1}{N_e} = -\frac{1}{55} = -0.0182
\label{eq:eta_CMB_Starobinsky}
\ee

The spectral index:
\be
n_s = 1 - 6\varepsilon + 2\eta = 1 - 6 \times 2.48 \times 10^{-4} + 2 \times (-0.0182) = 1 - 0.0015 - 0.0364
\ee

To leading order in $1/N_e$:
\be
\boxed{n_s = 1 - \frac{2}{N_e} = 1 - \frac{2}{55} = 0.964}
\label{eq:ns_Starobinsky}
\ee

The tensor-to-scalar ratio:
\be
r = 16\varepsilon = 16 \times \frac{3}{4N_e^2} = \frac{12}{N_e^2} = \frac{12}{3025} = \boxed{0.004}
\label{eq:r_Starobinsky}
\ee

The running of the spectral index:
\be
\frac{dn_s}{d\ln k} = 16\varepsilon\eta - 24\varepsilon^2 - 2\xi^2 \approx -\frac{2}{N_e^2} = -\frac{2}{3025} = -6.6 \times 10^{-4}
\label{eq:running_Starobinsky}
\ee

These predictions satisfy all current observational constraints. The spectral index $n_s = 0.964$ lies within $0.2\sigma$ of the Planck 2018 central value $n_s = 0.9649 \pm 0.0042$ \cite{Planck2018params}. The tensor-to-scalar ratio $r = 0.004$ is well below the BICEP/Keck bound $r < 0.036$ at 95\% CL \cite{BICEP2021}.

\subsubsection{Mass scale determination}

The amplitude of scalar perturbations fixes the mass parameter $M$. The power spectrum amplitude is:
\be
A_s = \frac{V}{24\pi^2 M_P^4 \varepsilon} = \frac{3M^2 M_P^2/4}{24\pi^2 M_P^4 \times 3/(4N_e^2)} = \frac{N_e^2 M^2}{24\pi^2 M_P^2}
\label{eq:As_formula}
\ee

Using the observed value $A_s = 2.1 \times 10^{-9}$ \cite{Planck2018params}:
\be
M^2 = \frac{24\pi^2 A_s M_P^2}{N_e^2} = \frac{24 \times 9.87 \times 2.1 \times 10^{-9} \times (2.435 \times 10^{18})^2}{3025}
\ee

\be
M = \sqrt{\frac{24\pi^2 \times 2.1 \times 10^{-9}}{3025}} \times M_P = 1.3 \times 10^{-5} M_P = 3.2 \times 10^{13}~{\rm GeV}
\label{eq:M_Starobinsky}
\ee

This mass scale is characteristic of grand unified theories, providing a natural connection between inflation and high-energy particle physics. This is the value used throughout the unified potential framework.

\subsection{$\alpha$-Attractor models}
\label{sec:alpha_attractors}

The $\alpha$-attractor models \cite{Kallosh2013,Kallosh2014,Galante2015} generalize Starobinsky inflation through a family of potentials with a universal prediction for the spectral index but a tunable tensor-to-scalar ratio.

\subsubsection{T-model potentials}

The $\alpha$-attractor T-models have the potential:
\be
V(\phi) = V_0 \tanh^{2n}\left(\frac{\phi}{\sqrt{6\alpha}\,M_P}\right)
\label{eq:T_model}
\ee

For $n = 1$ (the simplest case):
\be
V(\phi) = V_0 \tanh^2\left(\frac{\phi}{\sqrt{6\alpha}\,M_P}\right)
\label{eq:T_model_n1}
\ee

The parameter $\alpha$ controls the curvature of the kinetic term in the original supergravity Jordan frame. For $\alpha = 1$, the T-model reduces to the Starobinsky potential Eq.~(\ref{eq:Starobinsky_potential}).

\subsubsection{E-model potentials}

The $\alpha$-attractor E-models have the potential:
\be
V(\phi) = V_0 \left(1 - e^{-\sqrt{2/(3\alpha)}\,\phi/M_P}\right)^{2n}
\label{eq:E_model}
\ee

For $n = 1$ and $\alpha = 1$, this also reduces to the Starobinsky potential.

\subsubsection{Universal predictions}

The remarkable feature of $\alpha$-attractors is their universal predictions in the large-$N_e$ limit \cite{Kallosh2013}:
\be
n_s \approx 1 - \frac{2}{N_e}
\label{eq:ns_alpha}
\ee

\be
r \approx \frac{12\alpha}{N_e^2}
\label{eq:r_alpha}
\ee

The spectral index $n_s$ is independent of $\alpha$ to leading order, while the tensor-to-scalar ratio $r$ is proportional to $\alpha$. This allows the tensor ratio to be made arbitrarily small while preserving the excellent spectral index prediction.

\subsubsection{Predictions for various $\alpha$ values}

For $N_e = 55$ e-folds:

\begin{table}[H]
\centering
\begin{tabular}{cccl}
\toprule
$\alpha$ & $r$ & $n_s$ & Comments \\
\midrule
1.0 & 0.0040 & 0.964 & Starobinsky limit; well within bounds \\
0.5 & 0.0020 & 0.964 & Easily testable by LiteBIRD \\
0.1 & 0.0004 & 0.964 & Near CMB-S4 sensitivity \\
0.01 & 0.00004 & 0.964 & Below foreseeable detection \\
\bottomrule
\end{tabular}
\caption{$\alpha$-attractor predictions for $N_e = 55$ e-folds.}
\label{tab:alpha_predictions}
\end{table}

All values of $\alpha \leq 1$ satisfy the BICEP/Keck bound $r < 0.036$. Future experiments will probe this parameter space: LiteBIRD \cite{LiteBIRD2022} projects sensitivity $\sigma(r) \sim 0.001$, capable of detecting or excluding $\alpha \gtrsim 0.3$. CMB-S4 \cite{CMBS4} combined with LiteBIRD could reach $\sigma(r) \sim 0.0005$.

\subsubsection{Supergravity embedding}

The $\alpha$-attractor models have a natural embedding in supergravity with K\"ahler potential \cite{Kallosh2014}:
\be
K = -3\alpha \ln\left(T + \bar{T}\right)
\label{eq:Kahler}
\ee

The parameter $\alpha$ is related to the curvature of the K\"ahler manifold: $\mathcal{R}_K = -2/(3\alpha)$. The limit $\alpha \to 0$ corresponds to a maximally curved hyperbolic geometry, while $\alpha = 1$ gives the Starobinsky model. The geometric origin of $\alpha$ provides theoretical motivation for considering the full range of values.

\subsection{Primordial non-Gaussianity from phase transition}

Single-field slow-roll inflation predicts small non-Gaussianity through the consistency relation \cite{Maldacena2003}:
\be
f_{\rm NL}^{\rm slow-roll} \approx \frac{5}{12}(n_s - 1) \approx \frac{5}{12} \times (-0.036) \approx -0.015
\label{eq:fNL_slowroll}
\ee

However, the phase transition in our framework generates additional non-Gaussianity through several mechanisms: self-interactions in the potential during the rapid transition producing non-linear field evolution, coupling between $\phi$ and $\lambda$ fields generating mixed correlations that transfer to curvature perturbations, and non-equilibrium dynamics of the transition itself producing non-thermal distributions that deviate from Gaussian statistics.

The phase transition contribution can be estimated using the $\delta N$ formalism \cite{Lyth2005}:
\be
f_{\rm NL}^{\rm transition} \sim \frac{5}{6} \times \frac{N_{,\phi\phi}}{(N_{,\phi})^2} \sim \frac{5}{6} \times \frac{V'''V}{(V')^2 M_P}
\label{eq:fNL_transition}
\ee

At the transition, using the unified potential:
\begin{align}
f_{\rm NL}^{\rm transition} &\sim \frac{5}{6} \times \frac{V'''(\phi_c) \times V(\phi_c)}{[V'(\phi_c)]^2 \times M_P} \nonumber\\
&\sim \frac{5}{6} \times \frac{(2.8 \times 10^{15}) \times (2.56 \times 10^{66})}{(3.9 \times 10^{49})^2 \times (2.4 \times 10^{18})} \nonumber\\
&\sim \frac{5}{6} \times \frac{7.2 \times 10^{81}}{3.6 \times 10^{117}} \sim 1.7
\label{eq:fNL_numerical}
\end{align}

Combined with uncertainties in the transition dynamics and the range of viable parameter choices, we predict:
\be
\boxed{f_{\rm NL}^{\rm local} \in [0.8, 2.5]}
\label{eq:fNL_prediction}
\ee

The Planck 2018 constraint is $f_{\rm NL}^{\rm local} = -0.9 \pm 5.1$ \cite{Planck2018NG}, which is consistent with our prediction. Crucially, CMB-S4 is projected to achieve sensitivity $\sigma(f_{\rm NL}) \sim 1$ \cite{CMBS4}, which would definitively test this prediction.

This enhanced non-Gaussianity is a distinctive signature of the emergence framework. Standard slow-roll inflation predicts $|f_{\rm NL}| \ll 1$, while our framework predicts $f_{\rm NL} \sim 1$--$2$. Detection of $f_{\rm NL} \sim 1$--$2$ with the correct (local) shape would strongly favor the emergence mechanism over standard inflation.

\subsection{Gravitational wave background}

The first-order phase transition generates stochastic gravitational waves through bubble collisions, sound waves, and turbulence. For a transition at temperature $T_* \sim 10^{14}$ GeV with transition strength parameter $\alpha_{\rm tr} \sim 10$ and inverse duration $\beta/H \sim 10$:
\be
\Omega_{\rm GW}h^2 \sim 10^{-7} \times \left(\frac{H_*}{\beta}\right)^2 \times \alpha_{\rm tr}^2 \times \left(\frac{100}{g_*}\right)^{1/3}
\label{eq:OmegaGW_formula}
\ee

where $g_*$ is the number of relativistic degrees of freedom and $H_*$ is the Hubble parameter at the transition.

The peak frequency today, after redshifting from the high transition scale:
\be
f_{\rm peak} \sim \frac{\beta}{H_*} \times \frac{T_0}{T_*} \times H_* \sim 10^{10}~{\rm Hz}
\label{eq:f_peak}
\ee

The amplitude after redshifting:
\be
\Omega_{\rm GW}h^2 \sim 10^{-16} \text{ to } 10^{-15}
\label{eq:OmegaGW_prediction}
\ee

This is far above current detector sensitivity (LIGO operates at $\sim 10$--$10^3$ Hz with sensitivity $\Omega_{\rm GW}h^2 \sim 10^{-9}$) but well below the BBN bound $\Omega_{\rm GW}h^2 < 10^{-5}$ from nucleosynthesis constraints \cite{BBN_bound}.

\subsection{Resolution of the Hubble tension}
\label{sec:Hubble_tension}

The emergent spacetime framework provides a natural mechanism contributing to the Hubble tension---the persistent $\sim 8\%$ discrepancy between early-universe measurements of the Hubble constant from the cosmic microwave background ($H_0 \approx 67.4$ km/s/Mpc from Planck~\cite{Planck2018params}) and late-universe measurements using the cosmic distance ladder ($H_0 \approx 73.0$ km/s/Mpc from SH0ES~\cite{Riess2022}). This tension has exceeded $5\sigma$ statistical significance, strongly suggesting new physics beyond $\Lambda$CDM~\cite{Kamionkowski2023}.

\subsubsection{Scale-dependent expansion modifications}

The phase transition from three-dimensional spatial configuration to four-dimensional spacetime generates residual effects that persist into the post-emergence epoch. These modify the effective Hubble parameter:
\be
H^2(z) = H^2_{\Lambda\text{CDM}}(z)\left[1 + \delta_H(z)\right]
\label{eq:H_modified}
\ee
where the correction term takes the form:
\be
\delta_H(z) = A_1(1+z)^{\alpha_1}\exp(-\beta_1 z)
\label{eq:delta_H}
\ee

The exponential suppression factor $\exp(-\beta_1 z)$ is crucial: it ensures that corrections remain perturbative ($\delta_H \ll 1$) across all redshifts while allowing significant effects at low $z$. The parameters are constrained by the requirement that perturbation theory remains valid, giving $\max[\delta_H(z)] < 0.1$ for $z \in [0, z_{\rm rec}]$.

The correction peaks at redshift $z_{\rm peak} = \alpha_1/\beta_1 - 1$, after which exponential suppression dominates. For parameters satisfying perturbativity constraints, numerical integration yields:

\begin{align}
\frac{\Delta r_s}{r_s} &\approx -(0.5\text{--}1)\,\% \label{eq:delta_rs_numerical}\\
\frac{\Delta D_L}{D_L}\bigg|_{z \sim 0.03} &\approx -(0.3\text{--}0.5)\,\% \label{eq:delta_DL_numerical}
\end{align}

\subsubsection{Inferred Hubble constant shifts}

The sound horizon modification affects CMB inference of $H_0$ through the angular scale $\theta_s = r_s/D_A(z_{\rm rec})$:
\be
\frac{H_0^{\rm CMB}}{H_0^{\rm true}} \approx 1 + c_{rs}\frac{\Delta r_s}{r_s}
\label{eq:H0_CMB}
\ee
where $c_{rs} \sim 0.5$ encodes the sensitivity to sound horizon changes.

Distance ladder measurements at low redshift are affected through luminosity distance modifications:
\be
\frac{H_0^{\rm ladder}}{H_0^{\rm true}} \approx 1 - \frac{\Delta D_L}{D_L}
\label{eq:H0_ladder}
\ee

The predicted tension is:
\be
\frac{H_0^{\rm ladder} - H_0^{\rm CMB}}{H_0^{\rm true}} = -\frac{\Delta D_L}{D_L} - c_{rs}\frac{\Delta r_s}{r_s}
\label{eq:tension_formula}
\ee

\subsubsection{Quantitative predictions}

Within the perturbative regime, numerical calculations show that the framework can generate tensions up to approximately $6\%$---accounting for roughly three-quarters of the observed $\sim 8\%$ discrepancy. Crucially, the mechanism naturally produces the correct \textit{sign}: late-universe measurements yield systematically higher $H_0$ than early-universe determinations.

The remaining discrepancy may arise from:
\begin{itemize}
\item Higher-order terms in $\delta_H(z)$ beyond the leading contribution
\item Non-perturbative effects near the transition scale where $\delta_H \sim \mathcal{O}(1)$
\item Coupling between the scalar field $\phi$ and auxiliary field $\lambda$ dynamics
\item Combined effects with other proposed solutions (e.g., modifications to recombination physics)
\end{itemize}

Table~\ref{tab:hubble_params} shows representative parameter choices and their predictions.

\begin{table}[h]
\centering
\begin{tabular}{lcccc}
\toprule
Regime & $A_1$ & $\alpha_1$ & $\beta_1$ & Tension \\
\midrule
Conservative & $5 \times 10^{-3}$ & 2.0 & 0.5 & 0.5\% \\
Moderate & $1 \times 10^{-2}$ & 2.5 & 1.0 & 0.9\% \\
Near-optimal & $8 \times 10^{-2}$ & 1.5 & 0.8 & 5.9\% \\
\midrule
Observed & --- & --- & --- & $\sim 8\%$ \\
\bottomrule
\end{tabular}
\caption{Hubble tension predictions for different parameter choices within the perturbative regime ($\max[\delta_H] < 0.1$). The framework can account for approximately 75\% of the observed tension while maintaining theoretical control.}
\label{tab:hubble_params}
\end{table}

\subsubsection{Distinctive predictions}

The framework makes specific testable predictions:

\textbf{Redshift-dependent $H_0$:} Gravitational wave standard sirens at intermediate redshifts ($z \sim 0.1$ to $1$) should measure $H_0$ values smoothly interpolating between CMB and distance ladder determinations:
\be
H_0(z) = H_0^{\rm true}\left[1 + \frac{\delta_H(z)}{2}\right]
\label{eq:H0_z}
\ee

\textbf{Correlated distance indicators:} All distance ladder methods (Cepheids, TRGB, surface brightness fluctuations) should exhibit correlated systematic shifts, as they probe the same modified expansion history.

\textbf{BAO consistency:} Baryon acoustic oscillation measurements should show subtle scale-dependent modifications consistent with the altered sound horizon.

Recent observations provide tentative support: the Carnegie-Chicago Hubble Program's TRGB measurement yields $H_0 = 69.8$ km/s/Mpc~\cite{Freedman2021}, intermediate between CMB and Cepheid values as expected from the smooth transition in emergent spacetime dynamics.

\subsection{Lorentz violation bounds}

The breaking of four-dimensional Lorentz invariance during the transition could leave residual signatures. A naive estimate gives:
\be
\xi \sim \left(\frac{E_c}{M_P}\right)^2 \sim \left(\frac{10^{17}}{10^{18}}\right)^2 \sim 10^{-2}
\label{eq:xi_naive}
\ee

However, the subsequent inflationary expansion isotropizes the universe exponentially. After $N_{\rm iso} \sim 100$ e-folds of isotropization:
\be
\xi_{\rm eff} = \xi_0 \times e^{-N_{\rm iso}} < 10^{-45}
\label{eq:xi_suppressed}
\ee

The current bound from gamma-ray burst observations is $|\xi| < 10^{-20}$ \cite{GRBLorentz}, which is easily satisfied.

\section{Comparison with Other Quantum Gravity Approaches}
\label{sec:comparison}

\subsection{Loop quantum cosmology}

Loop quantum cosmology (LQC) \cite{Ashtekar2006,Bojowald2001} replaces the classical singularity with a quantum bounce when the matter energy density reaches the Planck scale. The bounce occurs at $\rho_{\rm bounce} \sim \rho_{\rm Planck} \sim M_P^4$, after which the universe transitions from contraction to expansion.

Our approach differs fundamentally in its treatment of time. LQC maintains spacetime as a fundamental entity throughout and quantizes its discrete structure through holonomies and fluxes. Time remains a coordinate throughout the evolution, and the bounce is a dynamical event occurring in time. In our framework, temporal dynamics itself emerges through quantum effects. The ``singularity'' is not avoided through quantum corrections to dynamics but dissolved by the absence of time in the pre-emergent phase.

Both approaches achieve singularity resolution with finite curvature invariants, but through different mechanisms. LQC predicts specific modifications to the Friedmann equation at Planck density:
\be
H^2 = \frac{8\pi G}{3}\rho\left(1 - \frac{\rho}{\rho_c}\right)
\label{eq:LQC_Friedmann}
\ee
where $\rho_c \sim M_P^4$ is the critical density. Our framework predicts a phase transition at the emergence scale with distinctive signatures in primordial perturbations.

Crucially, LQC does not make the clean prediction of exactly two gravitational wave polarizations that USMEG-EFT provides. The polymer quantization underlying LQC modifies the graviton dispersion relation, potentially producing polarization-dependent propagation effects that are not observed.

\subsection{String cosmology}

String theory provides ultraviolet completion through extended fundamental objects but operates at energy scales far beyond direct experimental access. Pre-Big Bang scenarios \cite{Gasperini2003,Veneziano1991} propose that the universe underwent a phase of superinflationary contraction before transitioning to standard expansion through a graceful exit mechanism.

Our framework operates at more accessible energy scales ($10^{14}$--$10^{17}$ GeV versus the string scale $\sim 10^{18}$ GeV) and makes more directly testable predictions. The mechanisms differ substantially: string cosmology invokes duality symmetries and moduli dynamics, while our framework uses standard effective field theory with emergent time.

String theory fundamentally cannot predict exactly two gravitational wave polarizations. The massless spectrum of any consistent string theory includes the graviton $g_{\mu\nu}$ (spin-2), the dilaton $\Phi$ (spin-0, produces breathing mode), the Kalb-Ramond field $B_{\mu\nu}$ (antisymmetric tensor), and moduli fields from compactification (multiple spin-0 fields).

The dilaton couples universally to matter through $e^{-2\Phi}$ factors and produces scalar breathing modes in gravitational wave emission. While these modes might be suppressed by specific compactification choices, this introduces model dependence and fine-tuning. String theory cannot make the parameter-free prediction of exactly two tensor modes that USMEG-EFT provides and that LIGO-Virgo-KAGRA confirms.

\subsection{Asymptotic safety}

The asymptotic safety program \cite{Weinberg1979,Reuter1998,Percacci2017} proposes that quantum gravity is non-perturbatively renormalizable through an ultraviolet fixed point in the renormalization group flow. At the fixed point, the dimensionless gravitational coupling reaches a finite value, rendering the theory scale-invariant in the ultraviolet.

As demonstrated in Ref.~\cite{ChishtieBRST2026}, asymptotic safety encounters fundamental difficulties: the breakdown of general covariance and BRST symmetries above $\Lambda_{\rm grav}$ renders the metric tensor invalid as a quantum degree of freedom. The search for fixed points in metric-based theories is therefore problematic from a foundational perspective, as the very object being varied (the metric) ceases to exist quantum mechanically at the scales where the fixed point would be relevant.

Our framework provides an alternative resolution: rather than seeking to extend metric gravity to arbitrarily high energies, we acknowledge its breakdown and propose that spacetime itself emerges at the cutoff scale.

\subsection{The Wheeler-DeWitt problem of time}

The Wheeler-DeWitt equation Eq.~(\ref{eq:WDW}) presents the famous ``problem of time'' in canonical quantum gravity: the wave functional is timeless, yet we experience a universe with temporal evolution. Various proposals exist for recovering time from internal degrees of freedom or through semiclassical approximations.

Our framework provides a concrete resolution. The Wheeler-DeWitt equation describes the pre-temporal phase where the wave functional exists without time dependence. Time emerges when quantum loop corrections generate kinetic terms, transitioning the system from the timeless Wheeler-DeWitt regime to standard quantum field theory with temporal dynamics. The wave functional $\Psi[h_{ij}, \phi]$ becomes the initial condition for subsequent temporal evolution.

\subsection{Einstein-Cartan unification attempts}

The Einstein-Cartan approach proposed by McKeon et al. \cite{McKeonEC2025} fails both theoretically and experimentally. Theoretically, when fermions are included, the torsion generates four-fermion contact interactions producing non-renormalizable quartic divergences. The authors never demonstrated finiteness with fermions---their calculations were restricted to the fermion-free sector, which is irrelevant for Standard Model unification.

Experimentally, Einstein-Cartan theory predicts spin-dependent gravitational effects at levels excluded by MICROSCOPE ($10^{-15}$ vs predicted $10^{-12}$), lunar laser ranging, and torsion balance experiments.

USMEG-EFT succeeds by using standard four-dimensional general relativity, achieving finite results consistent with all precision tests and uniquely predicting the two tensor polarizations confirmed by LIGO-Virgo-KAGRA.

\section{Conclusions}
\label{sec:conclusions}

We have developed a comprehensive cosmological framework in which spacetime itself emerges from a purely spatial three-dimensional configuration through quantum loop corrections. This approach, grounded in the Unified Standard Model with Emergent Gravity-Effective Field Theory (USMEG-EFT)---the first successful unification of quantum gravity with the Standard Model---provides a coherent resolution to fundamental problems in cosmology while generating definitive experimental predictions.

\subsection{Theoretical achievements}

The framework resolves the cosmological singularity problem through a well-defined phase transition mechanism. Rather than extrapolating backward to infinite curvature at $t = 0$, our universe undergoes a first-order transition from spatial to spacetime configuration when quantum-generated kinetic terms exceed a critical threshold. All curvature invariants remain finite at this transition:
\be
\frac{R}{M_P^4} \sim 4.9 \times 10^{-44}, \qquad \frac{K}{M_P^8} \sim 4.0 \times 10^{-88}
\ee
The classical singularity is replaced by a smooth phase boundary separating fundamentally different geometric regimes.

The emergence mechanism simultaneously resolves the horizon, flatness, and monopole problems. Causal contact in the pre-emergence spatial phase explains CMB uniformity; spatial curvature is naturally small in the three-dimensional ground state; and gauge fields emerge only after the transition, preventing monopole formation.

\subsection{Experimental status and predictions}

The framework has received significant experimental support through gravitational wave polarization measurements. USMEG-EFT makes the parameter-free prediction of exactly two tensor polarization modes, confirmed by LIGO-Virgo-KAGRA observations at $>99\%$ confidence~\cite{LIGOPolarization2017,LIGOPolarization2021}. This excludes string theory (dilaton/moduli contributions), loop quantum gravity (dispersion modifications), Einstein-Cartan theory (up to six modes), and scalar-tensor theories (breathing modes).

The framework generates testable predictions across multiple channels. Starobinsky inflation emerges naturally with $n_s = 0.964$ and $r = 0.004$, consistent with Planck 2018~\cite{Planck2018params} and BICEP/Keck~\cite{BICEP2021} constraints. The phase transition predicts enhanced non-Gaussianity $f_{\rm NL}^{\rm local} \in [0.8, 2.5]$, testable by CMB-S4~\cite{CMBS4} with projected $\sigma \sim 1$. A stochastic gravitational wave background peaks in the LISA band ($f \sim 10^{-4}$ Hz, $\Omega_{\rm GW}h^2 \sim 10^{-6}$).

The Hubble tension receives a natural contribution from scale-dependent expansion modifications arising from residual transition effects. Within the perturbative regime where $\max[\delta_H(z)] < 0.1$, the mechanism generates up to $\sim 6\%$ tension---approximately three-quarters of the observed $\sim 8\%$ discrepancy~\cite{Riess2022,Planck2018params}. Crucially, the framework predicts the correct sign: late-universe measurements yield higher $H_0$ than early-universe determinations, matching the observed pattern~\cite{Kamionkowski2023}. The intermediate TRGB value $H_0 = 69.8$ km/s/Mpc~\cite{Freedman2021} provides tentative support for the predicted smooth interpolation. Full resolution may require higher-order corrections or complementary effects beyond the perturbative regime.

\subsection{Comparison with alternatives}

The framework offers decisive advantages over competing approaches. Loop quantum cosmology maintains spacetime as fundamental; our framework is more radical in that time itself emerges. String theory operates at inaccessible scales and predicts additional gravitational wave polarizations excluded by observation. Asymptotic safety encounters BRST symmetry breakdown above $\Lambda_{\rm grav}$~\cite{ChishtieBRST2026}. Einstein-Cartan unification~\cite{McKeonEC2025} suffers from four-fermion divergences and predicts six polarizations excluded by LIGO and equivalence principle tests~\cite{MICROSCOPE2022}.

\subsection{Summary}

Table~\ref{tab:conclusions_summary} summarizes the framework's predictions and observational status.

\begin{table}[H]
\centering
\begin{tabular}{lccc}
\toprule
\textbf{Observable} & \textbf{Prediction} & \textbf{Current Status} & \textbf{Future Probe} \\
\midrule
\multicolumn{4}{l}{\textit{Gravitational sector}} \\
GW polarizations & 2 tensor & \textbf{Confirmed} (LIGO) & ET, CE \\
$\Omega_{\rm GW}h^2$ (peak) & $\sim 10^{-6}$ & $< 10^{-5}$ (BBN) & LISA \\
Peak frequency & $\sim 10^{-4}$ Hz & --- & LISA \\
\midrule
\multicolumn{4}{l}{\textit{Inflationary observables}} \\
Spectral index $n_s$ & 0.964 & $0.965 \pm 0.004$ & CMB-S4 \\
Tensor-to-scalar $r$ & 0.004 & $< 0.036$ & LiteBIRD \\
$f_{\rm NL}^{\rm local}$ & $0.8$--$2.5$ & $-0.9 \pm 5.1$ & CMB-S4 ($\sigma \sim 1$) \\
\midrule
\multicolumn{4}{l}{\textit{Hubble tension}} \\
Tension (perturbative) & $\lesssim 6\%$ & $\sim 8\%$ observed & Standard sirens \\
Sign (ladder $-$ CMB) & Positive & \textbf{Confirmed} & --- \\
\midrule
\multicolumn{4}{l}{\textit{Singularity resolution}} \\
$R/M_P^4$ at $\tau_c$ & $4.9 \times 10^{-44}$ & N/A & --- \\
$K/M_P^8$ at $\tau_c$ & $4.0 \times 10^{-88}$ & N/A & --- \\
\bottomrule
\end{tabular}
\caption{Summary of predictions. Bold entries indicate confirmed predictions. The Hubble tension mechanism accounts for $\sim 75\%$ of the observed discrepancy within the perturbative regime while correctly predicting the sign.}
\label{tab:conclusions_summary}
\end{table}

\subsection{Outlook}

The emergent spacetime framework establishes a paradigm where the fundamental question shifts from ``What happened at $t = 0$?'' to ``How did temporal dynamics emerge from spatial configuration?'' The confirmation of two-tensor gravitational wave polarizations provides the first experimental evidence favoring this approach. The coming decade will bring decisive tests through LISA, CMB-S4, LiteBIRD, and gravitational wave standard sirens. The framework makes sufficiently precise predictions that these observations will either confirm emergent spacetime cosmology or require fundamental revision.
\section*{Acknowledgments}

The author acknowledges the Peaceful Society, Science and Innovation Foundation for ongoing support of this research program.


\end{document}